\documentclass{trbunofficial}
\usepackage{graphicx}

\usepackage{algorithm}
\usepackage{tabularx}
\usepackage{algpseudocode}
\usepackage{subcaption}

\usepackage[hidelinks]{hyperref}

\AuthorHeaders{Jiang, Zhou, Y. Zhao et al.}
\title{From Patchwork to Network: A Comprehensive Framework for Demand Analysis and Fleet Optimization of Urban Air Mobility
}

\author{%
  \textbf{Xuan Jiang, Corresponding Author}\\
  Department of Urban Studies and Planning\\
  Massachusetts Institute of Technology, Cambridge, MA 02139\\
  Email: xuanj@mit.edu\\
  \hfill\break
  \textbf{Xuanyu Zhou}\\
  College of Civil Engineering and Architecture\\
  Zhejiang University, Hangzhou, China 310058\\
  Email: xuanyu.zhou@zju.edu.cn\\
  \hfill\break
  \textbf{Yibo Zhao}\\
  Department of Civil \& Systems Engineering\\
  Johns Hopkins University, Baltimore, MD 21218\\
  Email: yzhao231@jh.edu\\
  \hfill\break
  \textbf{Shangqing Cao}\\
  Department of Civil and Environmental Engineering\\
  University of California, Berkeley, Berkeley, CA 94720\\
  Email: caoalbert@berkeley.edu\\
  \hfill\break
  \textbf{Haoze He}\\
  School of Computer Science\\
  Carnegie Mellon University, Pittsburgh, PA 15213\\
  Email:haozeh@cs.cmu.edu\\
  \hfill\break
  \textbf{Jinhua Zhao, Ph.D.}\\
  Department of Urban Studies and Planning\\
  Massachusetts Institute of Technology, Cambridge, MA 02139\\
  Email:jinhua@mit.edu\\
  \hfill\break%
    \textbf{Mark Hansen, Ph.D.}\\
  Department of Civil and Environmental Engineering\\
  University of California, Berkeley, Berkeley, CA 94720\\
  Email: hansen2@berkeley.edu \\
  \hfill\break%
  \textbf{Raja Sengupta, Ph.D.}\\
  Department of Civil and Environmental Engineering\\
  University of California, Berkeley, Berkeley, CA 94720\\
  Email: rajasengupta@berkeley.edu
}

\TotalWords{4699}

\begin{document}
\maketitle

\section*{Abstract}
Urban Air Mobility (UAM) presents a transformative vision for metropolitan transportation, but its practical implementation is hindered by substantial infrastructure costs and operational complexities. We address these challenges by modeling a UAM network that leverages existing regional airports and operates with an optimized, heterogeneous fleet of aircraft. We introduce LPSim, a Large-Scale Parallel Simulation framework that utilizes multi-GPU computing to co-optimize UAM demand, fleet operations, and ground transportation interactions simultaneously. 

Our equilibrium search algorithm is extended to accurately forecast demand and determine the most efficient fleet composition. Applied to a case study of the San Francisco Bay Area, our results demonstrate that this UAM model can yield over 20 minutes' travel time savings for 230,000 selected trips. However, the analysis also reveals that system-wide success is critically dependent on seamless integration with ground access and dynamic scheduling.

\hfill\break%
\noindent\textit{Keywords}: Urban Air Mobility, UAM fleet operation, Regional-scale traffic simulation, Heterogeneous fleet, Transportation network equilibrium
\newpage

\section{Introduction}
Urban transportation systems around the world face unprecedented strain due to the increase in population density and the expansion of the urban landscape \cite{hae_choi_exploring_2022}. Recent studies indicate that approximately 56 58\% of the world's population live in urban areas \cite{li_rethinking_2024,world_bank_urban_2024}, leading to severe traffic congestion, economic losses, and environmental degradation, particularly in megacities such as New York, Los Angeles and Tokyo \cite{qu_demand_2024}. Addressing these challenges requires prioritizing strategic planning of transportation networks and requires innovative approaches to urban mobility.

In response, urban air mobility (UAM) has emerged as a transformative paradigm, proposing the use of low-altitude airspace for on-demand passenger transport \cite{cohen_urban_2021}. The development of electric vertical take-off and landing (eVTOL) aircraft by pioneering companies such as Joby Aviation, EHang, Archer Aviation, and Volocopter is a critical enabler of this vision, promising a safer, quieter, and more sustainable alternative to ground-based travel \cite{long_demand_2023}. However, the immense capital investment required to build a new network from the ground up presents a significant barrier \cite{hae_choi_exploring_2022, ryerson_build_2014}. This challenge highlights the strategic importance of existing aviation infrastructure. For example, the New York and Los Angeles metropolitan areas contain 25 and 24 regional airports, respectively, many with surplus capacity \cite{bonnefoy_emergence_2006, ryerson_build_2014}. Therefore, re-purposing existing and underutilized regional airports offers a viable entry point for initiating UAM operations.

However, transitioning UAM from a promising concept to a viable reality presents immense system-level challenges beyond infrastructures. Successful launch and operation of an UAM service requires solving a series of additional problems, including demand modeling and estimation, fleet sizing, network and corridor design, and real-time fleet management under uncertainty\cite{long_demand_2023}. Among these, heterogeneous fleet composition emerges as a critical factor, as it directly influences capital investment and operating efficiency \cite{kim_receding_2020, cao2024fleetsizespilluam}. Optimizing aircraft selection to match spatiotemporal demand variations minimizes seat wastage while ensuring service coverage, thereby crucially influencing system viability. 

Furthermore, conventional transportation analysis tools are often inadequate for this task \cite{koliogeorgi_auto-tuning_2024}. They struggle with the scale of regional analysis and often lack the ability to model the dynamic feedback loop between emerging air traffic and existing ground transportation systems. The significant computational time required by conventional simulators creates a performance bottleneck, hindering the large-scale, iterative analysis needed for robust system design and optimization \cite{jiang_large_2024, saprykin_gemsim_2019}.

In this article, we present a comprehensive tool-based framework for systematic planning, control, and analysis of UAM systems. The proposed framework is designed to address two primary research questions: (1) What is the maximum demand that a UAM system can serve by fully leveraging the existing regional airport infrastructure? (2) How can a hypothetical fleet of UAM vehicles be optimally constructed to accommodate the diurnal variations in demand?

The remainder of this paper is organized as follows. Section \ref{sec:Related_works} reviews the existing literature on UAM demand calculation and UAM operation simulation, including an introduction to our previous contributions. Section \ref{sec:Methodology} details the methodology and computational architecture employed for the allocation of UAM ground transportation and the sizing of heterogeneous fleet. Section \ref{sec:Case_Study} presents the specifics of our San Francisco Bay Area simulation case, covering data sources, simulation setup, and calculation bases. Section \ref{sec:Result} discusses our primary findings from the simulation and compares them with existing literature. Finally, Section \ref{sec:Conclusion} concludes the paper and outlines the directions for future research. 

\section{Related Works}
\label{sec:Related_works}
\subsection{UAM Demand Calculation}
Strategic network planning is the cornerstone of successful implementation of UAM, and it fundamentally relies on accurate travel demand forecasting \cite{cohen_urban_2021, long_demand_2023, rimjha_urban_2021}. This section reviews existing literature on UAM demand estimation to highlight a critical methodological limitation: the predominant use of static models that neglect the dynamic feedback between air and ground transportation systems.

Current research primarily estimates UAM demand by modeling mode choice based on the attributes of trips across different transportation options. In this framework, a trip qualifies for UAM if it provides substantial savings in travel time or cost relative to ground transportation \cite{kim_potential_2023, hae_choi_exploring_2022, peng_hierarchical_2022} Some studies also suggest that airport access and egress times influence demand \cite{coppola_urban_2025, pukhova_flying_2021, rimjha_urban_2021}. Researchers mainly draw on static data to perform demand analysis, including current airport location \cite{rimjha_urban_2021, rimjha_urban_2021-1}, Uber trip data \cite{rimjha_urban_2021-1}, census-level socioeconomic data \cite{haan_are_2021, straubinger_proposing_2021}, cell phone data\cite{haan_are_2021}, and ground-level transportation data \cite{bulusu_traffic_2021, kim_potential_2023}. Different application of UAM also impact the forecasting of demand, a study in Milan found UAM airport shuttles were more attractive than on-demand air taxis, capturing a potential modal share of 2-5\% for airport trips, compared to 1-3\% for air taxis \cite{coppola_urban_2025}. 

The critical limitation of this static approach is its inability to account for the dynamic feedback loop between UAM and ground transportation systems \cite{jiang2024simulatingintegrationurbanair}. As UAM adoption increases, it diverts traffic from terrestrial routes, which can partially alleviate road congestion. By analyzing demand based on a fixed, pre-existing state of congestion, static models overlook this shifting equilibrium. As a result, they are likely to produce an overestimation of the long-term sustainable demand for UAM services. Future research must therefore evolve towards dynamic models that can capture the reciprocal effects between air and ground mobility to generate more realistic and reliable demand forecasts \cite{kim_receding_2020}.

\subsection{UAM Operation Simulator}
The operational management of UAM is exceptionally complex, involving dynamic passenger demand, intricate airspace interactions, and energy-constrained vehicle fleets \cite{song_development_2021, thu_multivehicle_2022, 10637246}. Due to these interconnected complexities, analytical solutions are often intractable, making high-fidelity simulation platforms indispensable as virtual testbeds for developing and validating operational algorithms \cite{chen_integrated_2024, lee_functional_2024}. 

These simulators provide a dynamic environment to assess the performance and robustness of various management strategies before real-world deployment. For instance, in the domain of on-demand fleet scheduling, simulators are used to evaluate the effectiveness of optimization methods like Mixed-Integer Linear Programming (MILP) and imitation learning in minimizing passenger wait times \cite{kleinbekman_rolling-horizon_2020, poddar_graph-based_2024, wu_optimal_2020}. Simulators also provide the necessary environment to assess the practical viability and effectiveness of heuristic approaches, such as Genetic Algorithms (GA) and Particle Swarm Optimization (PSO), for managing large-scale, heterogeneous fleets \cite{kim_receding_2020}. 

Furthermore, simulators are critical for validating advanced airspace management strategies. To ensure safety and efficiency in high-density airspace, Demand-Capacity Balancing (DCB) has emerged as another focal point, with solutions relying on advanced AI techniques like Multi-Agent Reinforcement Learning (MARL) for strategic traffic deconfliction \cite{huang_strategic_2022, murthy_reinforcement_2025, xie_reinforcement_2021} and AI-driven urban airspace monitoring method \cite{liu_intelligent_2025}. At the level of real-time dispatching, VertiSync, a centralized, conflict-free takeoff scheduling policy that explicitly integrates both trip request servicing and critical vehicle rebalancing within UAM networks \cite{pooladsanj_throughput_2025}. Building on this, our previous research introduced a Multi-modal Regional-scale Traffic Simulation (MRTS), which provides an integrated environment to model the coupled dynamics of passengers, aircraft, and ground traffic, enabling more holistic algorithm validation \cite{jiang2024simulatingintegrationurbanair, cao2024fleetsizespilluam}. 

While these operational algorithms are powerful in theory, their practical development and large-scale validation are constrained by the performance of the simulation environments \cite{koliogeorgi_auto-tuning_2024,saprykin_application_2022, saprykin_gemsim_2019}. For any operational strategy, a simulator must validate thousands of diverse, large-scale vehicle scenarios. As a result, the simulator’s speed directly determines the comprehensiveness and efficiency of this validation process \cite{aqib_rapid_2019, saprykin_accelerating_2022}. Further more, this constraint is even more acute for AI-driven methods like reinforcement learning, where the simulator is not just a validation tool but an integral part of the training loop \cite{murthy_reinforcement_2025}. 

Presently, the vast majority of UAM simulation platforms are built on CPU-based architectures. This creates a computational bottleneck, rendering large-scale, real-time simulations computationally prohibitive and thus severely limiting the scope and speed of algorithmic validation \cite{cao_integrating_2024}. Therefore, a critical research gap exists between the needs of advanced algorithms and the capabilities of current simulators. Our previous study has developed LPSim \cite{jiang_large_2024, jiang2024designing}, a Large-Scale Multi-GPU Parallel Computing-based Regional Scale Traffic Simulation Framework, which not help boost the computation process but also maintain the accuracy with its careful calibration \cite{jiang2024drbo, jiang2025drbo}.  

\section{Methodology}
\label{sec:Methodology}
\subsection{UAM and Ground Trip Simulation}
Our simulation's primary objective is to seamlessly integrate UAM into existing transportation systems as a multimodal mode. This integration encompasses the entire journey, from first-mile to last-mile ground segments to the air trips. Consequently, our framework explicitly captures the intricate interplay between UAM operations and traditional ground transportation \cite{jiang_large_2024, jiang2025simulation}.

\begin{figure}[htbp]
  \centering
  \includegraphics[width=0.6\textwidth]{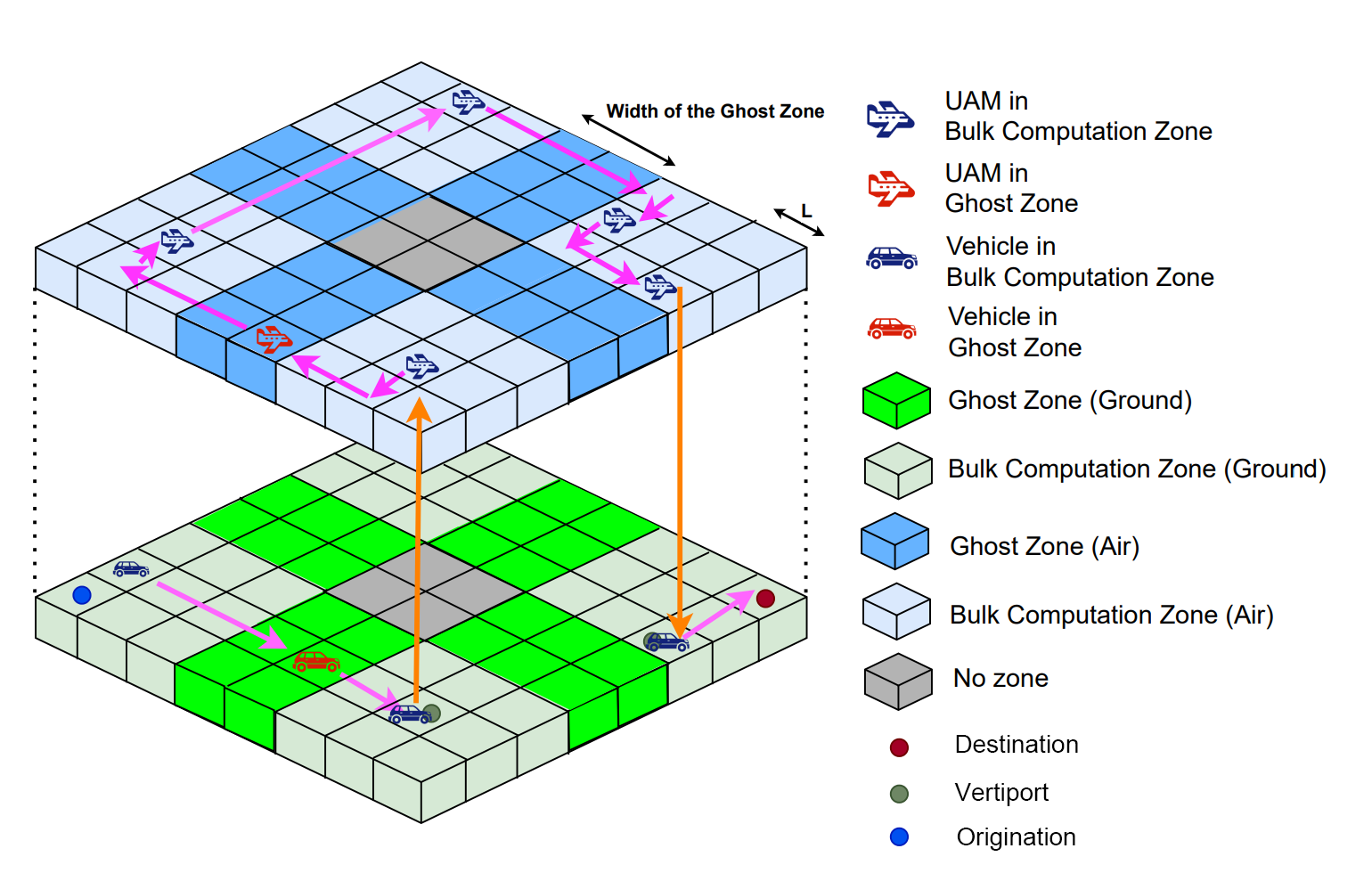}
  \caption{Illustration of the multi-layer road network for UAM simulation. UAM and ground vehicles move within designed areas on separate layers, maintaining independent computation and management. }\label{fig:two_layer}
\end{figure}

Our simulation begins with the generation of demand and routing stages, assigning a mode (Auto or UAM) to each segment of the trip. Ground transportation follows traditional traffic simulation paradigms, including car-following, lane-changing, and gap-acceptance models. For UAM trips, our system dynamically manages operations by continuously monitoring vertiport capacity and enforcing departure intervals to satisfy necessary operational separation requirements. The complete multimodal network, illustrating the distinct layers for UAM and ground traffic, is presented in Figure \ref{fig:two_layer}. 

As shown in Algorithm \ref{alg:uam_movement}, the input data for our simulation includes regional transportation networks, airport infrastructures, and demand data. Outputs comprise key performance indicators such as the total addressable market of UAM, operational benefits, and aircraft utilization rates. This methodology provides a robust framework for evaluating the integration of UAM into urban transportation networks, laying the groundwork for scalable UAM deployment and enhanced operational tools.

\begin{algorithm}[htbp]
\caption{UAM Simulation Process}
\label{alg:uam_movement}
\textbf{State}: $\mathcal{D}$ (Demand), $\mathcal{M}$ (Mode Identifier), $\mathcal{P}$ (Person), $\mathcal{V}$ (Airport Capacity), $\mathcal{C}$ (Current UAM Count), $\mathcal{E}$ (Simulation End Status)\\
\textbf{Parameter}: $T$ (Operation Separation Interval for Runway), $x$ (Departed Aircraft Count)\\
\textbf{Output}:  Passenger Travel Time
\begin{algorithmic}[1]
\While{Simulation is not ended}
    \State Generate demand
    \State Route each person through added mode identifier
    \If{Person mode is Auto}
        \State Simulate car following, lane changing, and gap acceptance as detailed in \cite{jiang2024lpsim}
    \ElsIf{Person mode is UAM}
        \If{Departed $x$ aircraft}
            \State Decrease UAM count by $x$
        \EndIf
        \If{$\mathcal{C}$ in Current Airport $\leq \mathcal{V}$}
            \State Depart person after $T$ interval
        \Else
            \State Hold person for $T$ interval
        \EndIf
    \EndIf
    \State Simulate until the next edge
    \If{End of Simulation}
        \State Exit the simulation loop
    \EndIf
\EndWhile
\State \textbf{return}  Passenger Travel Time
\end{algorithmic}
\end{algorithm}

\subsection{UAM Allocation Equilibrium Algorithm}

This subsection introduces the UAM Trip Allocation Equilibrium Algorithm designed to optimally and dynamically allocate trips between UAM and ground transportation modes within the LPSim system as shown in Algorithm \ref{alg:uam_allocation_equilibrium}. This algorithm simulates the day-to-day mode choice behavior of travelers, allowing the system to converge to a stable state where no traveler can unilaterally improve their travel time by switching modes.

\begin{algorithm}[tbp]
\small
\caption{UAM Trip Allocation Equilibrium Algorithm}
\label{alg:uam_allocation_equilibrium}
\textbf{State}: $\mathcal{T}_{OD}$ (Origin-Destination trips from SFCTA), $\mathcal{T}_{UAM}$ (UAM trips), $\mathcal{T}_{ground}$ (Ground trips), $t_{UAM}$ (UAM travel time), $t_{ground}$ (Ground travel time), $t_{driving}$ (Driving time threshold)\\
\textbf{Output}: Equilibrium allocation of UAM and ground trips, total addressable market for UAM, benefit/loss for each mode

\begin{algorithmic}[1]
\State Initialize $\mathcal{T}_{OD}$ with 6.6M trips representing morning peak demand in the Bay Area
\State Use existing regional airports for UAM allocation
\While{Equilibrium not reached}
    \State \textbf{Step 1: UAM Allocation}
    \State Allocate trips to UAM where $t_{UAM} \leq t_{driving}$
    \State Remove non-benefited UAM trips where $t_{UAM} > t_{driving}$ and reassign them to ground trips
    \State Update $t_{UAM}$ based on traffic simulation results from LPSim-UAM

    \State \textbf{Step 2: Ground Trip Reallocation}
    \State Reallocate ground trips to UAM where $t_{ground} > t_{driving}$ if it benefits from UAM travel time
    \State Update $t_{ground}$ based on traffic simulation results from LPSim-UAM

    \State \textbf{Convergence Check}
    \If{No significant changes in allocation between UAM and ground}
        \State Convergence achieved; exit loop
    \EndIf
\EndWhile

\State Calculate total addressable market for UAM
\State Calculate net benefit to UAM trips and benefit/loss to ground trips
\State \textbf{return} Equilibrium allocations, total addressable market, benefits and losses
\end{algorithmic}
\end{algorithm}

\subsubsection{Input Data and Initialization}

The process begins with an initial partition of the total OD trips, $\mathcal{T}_{OD}$, into preliminary sets of UAM trips $\mathcal{T}_{UAM}^{(0)}$ and ground trips $\mathcal{T}_{ground}^{(0)}$. 

The core of the algorithm is an iterative loop that continues until convergence is achieved. Each iteration, indexed by n, represents a cycle of mode shifting based on updated travel times. Therefore, our algorithm can model the dynamic feedback between the two transportation systems. The travel time for both UAM $t_{UAM}^{(n)}$ and ground transport $t_{ground}^{(n)}$ are recalculated at the beginning of each relevant step using our simulator.

\subsubsection{Iteration Steps}

Within each iteration, two primary reallocation steps occur. First, a "UAM Demotion" step evaluates the trips currently assigned to UAM. And any trip $(i,j)$ whose  UAM travel time $t_{UAM}^{(n)}(i,j)$ is found to be longer than the baseline pure driving time $t_{driving}$ is deemed non-beneficial. These travelers are assumed to switch back to ground transport, thus alleviating potential airspace or vertiport congestion. 

Second, a "Ground Promotion" step identifies potential candidates to switch from ground to air. Any traveler whose ground journey $t_{ground}^{(n)}(i,j)$ is longer than the baseline driving time is considered eligible for UAM. To ensure stable and gradual convergence, particularly in the initial stages, we promote a fraction $\beta$ of these eligible ground trips to the UAM set. This promotional mechanism is typically active for a  limited number of early iterations, controlled by the parameter $index_{threshold}$, after which travelers switch based on other emergent properties of the system.

This iterative process of demotion and promotion continues until the system stabilizes. Convergence is reached when the net change in the set of UAM users between two consecutive iterations becomes negligible. Specifically, the loop terminates when the proportion of trips that switched modes relative to the total number of OD trips falls below a predefined tolerance threshold $\epsilon$. Upon convergence, the algorithm outputs the final equilibrium allocations of $\mathcal{T}_{UAM}$ and $\mathcal{T}_{ground}$. This result should represent the total addressable market for UAM under realistic, congested conditions, along with the calculated net travel time benefits or losses for travelers in each mode.

\subsection{Fleet Size Optimization}
This section details a two-phase methodology to determine the optimal composition of a heterogeneous aircraft fleet. 

\subsubsection{Phase 1: Demand Characterization via Unconstrained Simulation}

In this initial phase, we operate under the assumption of an unconstrained or "infinite" fleet, meaning all aircraft types are available on demand. For each origin-destination (OD) trip qualifying for UAM, the LPSim model dispatches the most suitable aircraft type. Suitability is determined through a two-step process: first, filtering for aircraft types with an operational range sufficient for the trip distance and meet the runway requirement for the airport, and second, selecting the aircraft with the smallest seating capacity that is greater than or equal to the number of passengers in the travel party. 

This process transforms the final UAM trip $\mathcal{T}_{UAM}$ set obtained from Algorithm \ref{alg:uam_allocation_equilibrium} into a detailed flight task collection $T$. Each task in $T$ is defined by its origin, destination, operation time, and an ideal assigned aircraft type.

\subsubsection{Phase 2: Fleet Sizing via Network Flow Model}

With the targeted flight task collection $T$ established, we then apply a network flow model to calculate the minimum required fleet size. To determine the minimum fleet size required to service a given set of passenger tasks $T$, we formulate the problem as a minimum cost network flow model on a time-expanded graph $G(W,E)$.

The set of nodes $W$ consists of a unique start node $t_{start}$ and end node $t_{end}$ for each task $t \in T$, in addition to a global source and sink node:

$$
    W = \{t_{start} | t\in T\} \cup \{t_{end} | t\in T\} \cup \{source, sink\}
$$

The set of edges $E$ represents all valid movements and connections in the network. It includes:(1) task edges $(t_{start},t_{end}$ for each task $t \in T$, representing the completion of that trip; (2) source and sink edges connecting the global source to every $t_{start}$ and every $t_{end}$ to the global sink; and (3) transition edges $(i_{end},j_{start}$ that link the completion of task $i$ to the start of a subsequent task $j$: 

$$
    E = \{(t_{start}, t_{end}) | t\in T\} \cup \{(source, t_{start}) | t\in T\} \cup \{(t_{end}, sink) | t \in T\} \cup \{(i,j) | i_{end} + \Delta t_{ij} \leq  j_{start}\ i \in T, j \in T\}
$$

A transition edge exists only if a single aircraft can feasibly service both tasks, which means the condition $i_{end} + \Delta t_{ij} \leq j_{start}$ is met. Here, $\Delta t_{ij} = f_{ij} + c_{i}$ represents the total turnaround time, composed of the repositioning flight time $f_{ij}$ and necessary charging time required after completing task $i$.

The optimization model seeks to find the minimum flow from the source that can satisfy all tasks. The flow on each edge $(i,j)$ is denoted by the decision variable $x_{ij}$. To minimize the fleet size, we set the unit cost $c_{ij}$ to 1 for all edges originating from the source node, and to 0 for all other edges. Consequently, the objective function shown below is to minimize the total cost, where the resulting total flow from the source corresponds to the minimum number of aircraft required.

$$
\underset{x_{ij}}{min} \sum_{i} \sum_{j} x_{ij}c_{ij}
$$

subject to:

\setcounter{equation}{0}

\begin{align}
    x_{ij} \leq u_{ij} \quad \forall (i,j) \in E
    \label{equal:upper_bound}
\end{align}

\begin{align}
    x_{ij} \geq 0 \quad  \forall (i,j) \in E
    \label{equal:lower_bound}
\end{align}

\begin{align}
    \sum_{j} x_{ji} - \sum_{j} x_{ij} = n_{i} \quad \forall i \in W
    \label{equal:flow_conservation}
\end{align}

\begin{align}
    x_{ij} = 1 \quad \forall (i,j) \in \{(t_{start}, t_{end})\ |t \in T\}
    \label{equal:demand}
\end{align}

This minimization is subject to the following constraints. First, standard flow conservation must be maintained at every node $i \in W$ (Equation \ref{equal:flow_conservation}). Let $n_{i}$ be the node capacity of node $i$, $i \in W$. Let $n_{source} = -N$ and $n_{sink} = N$, where $N$ is a sufficiently large number that can serve the demand. All other nodes having a net flow of zero. Second, to ensure every passenger trip is serviced, the flow on each task edge $t_{start}, t_{end}$ is constrained to be exactly 1 (Equation \ref{equal:demand}). Finally, the flow on any edge must be non-negative (Equation \ref{equal:lower_bound}) and cannot exceed its capacity $u_{ij}$ (Equation \ref{equal:upper_bound}). For this problem, all edge capacities can be set to 1, as each connection represents the path of single aircraft. 

Based on this framework, we calculate the minimum fleet size for homogeneous fleets separately. For each aircraft type $k$, we extract its assigned tasks from the LPsim output collection $T$ and solve the optimization problem. This determines the minimum number of aircraft required if the fleet consisted solely of type $k$. Subsequently, we extend the model to consider all aircraft types simultaneously to ascertain the minimum required fleet for the specific composition proposed by LPSim.

\section{Case Study}
\label{sec:Case_Study}
\begin{figure}[!ht]
  \centering
  \includegraphics[width=0.6\textwidth]{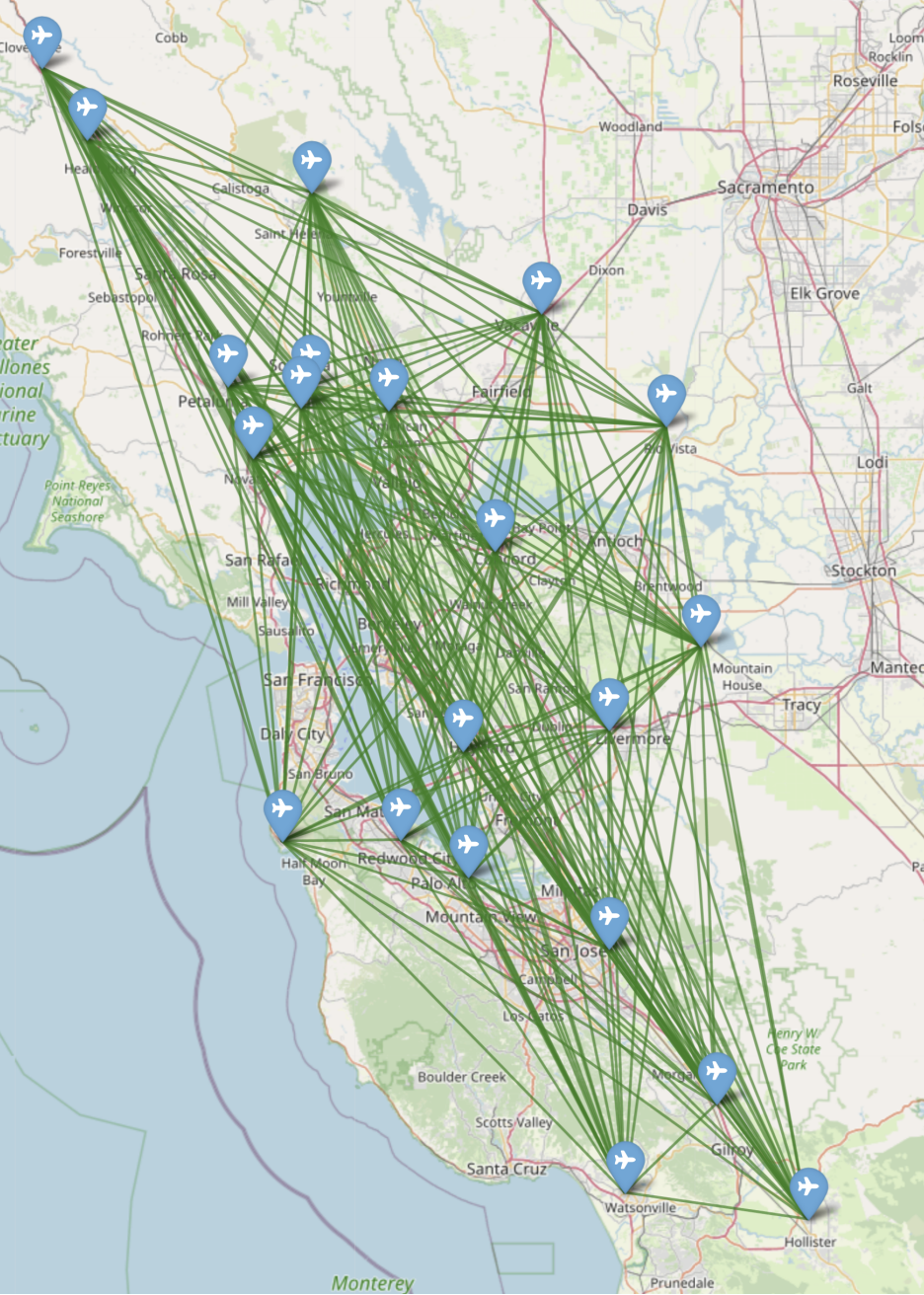}
  \caption{Distribution of 21 regional airports in San Francisco Bay Area}\label{fig:airport_distribution}
\end{figure}

This section presents a case study conducted in California to evaluate the framework's application. Our study assesses the potential utility of a large-scale regional network in the San Francisco Bay Area by evaluating a scenario where all 21 regional airports (Figure \ref{fig:airport_distribution}) are fully dedicated to UAM operations.

\subsection{Airports and Ground Transportation Data}
The road network for this study was sourced from OpenStreetMap (OSM), encompassing the San Francisco Bay Area with 223,327 nodes (intersections) and 547,696 edges (streets). 

Urban travel demand data was obtained from the San Francisco Chained Activity Modeling Process-6 (SF-CHAMP) framework, an advanced model developed by the San Francisco County Transportation Authority (SFCTA) that predicts transportation patterns based on infrastructure, travel behaviors, and socio-economic data \cite{outwater_san_2008,sall_evaluating_2010}. The dataset reflects a typical weekday, containing approximately 17.8 million trips over a full day and 6.6 million during the morning period (12:00 AM–12:00 PM), with each trip record detailing attributes such as origin, destination, and transportation mode \cite{chan_simulating_2023}. Finally, the airport network consists of 21 general aviation airports in the Bay Area, with detailed specifications for each provided in Table \ref{tab:airports}.

\begin{figure}[!ht]
  \centering
  \includegraphics[width=1\textwidth]{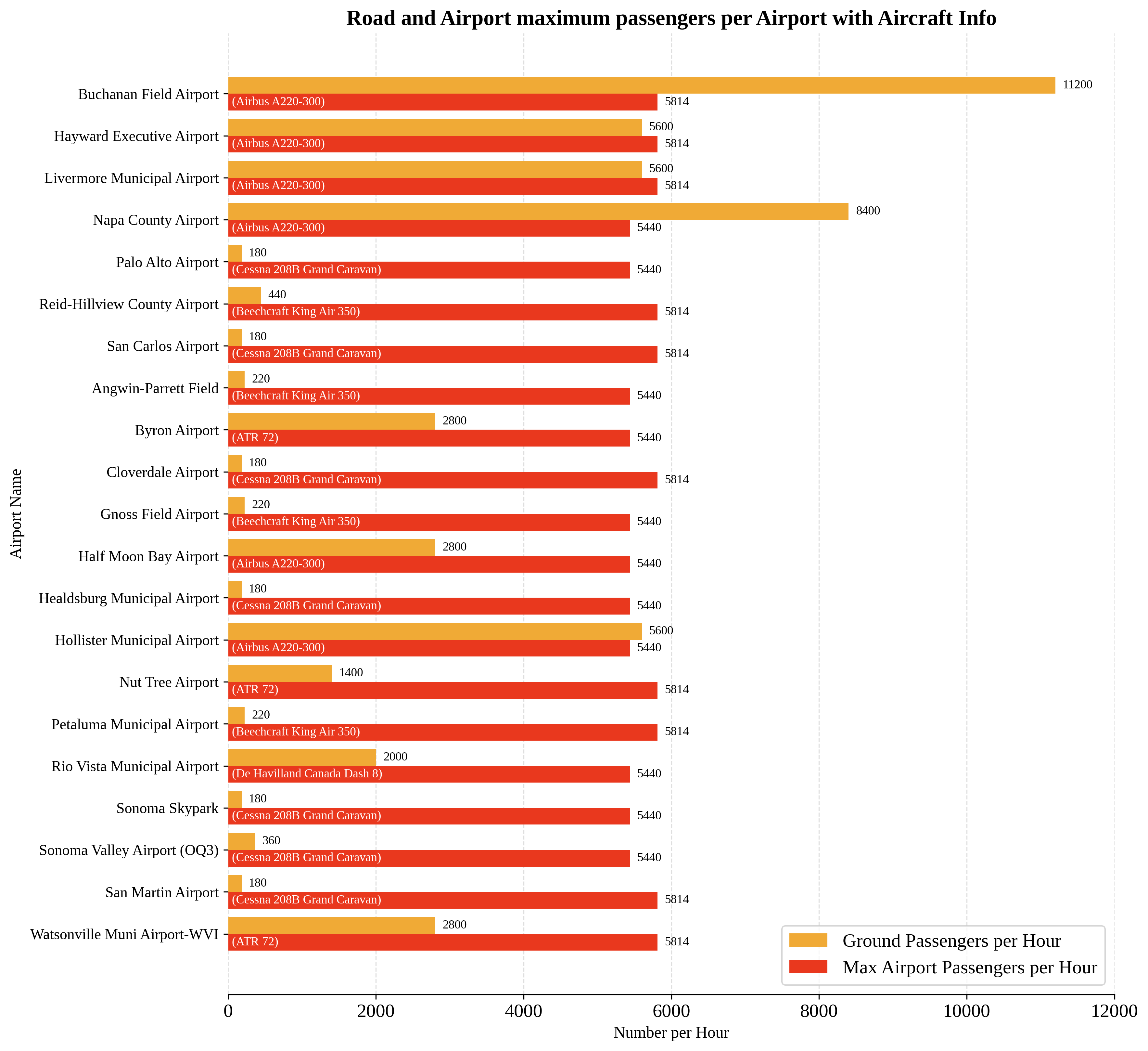}
  \caption{Road and Airport maximum passengers per Airport with Aircraft Info}\label{fig:road_airport_capacity}
\end{figure}

To estimate the potential capacity of each airport in the Bay Area, we evaluated both its ground-side and air-side limits. The maximum ground transportation capacity was calculated using an average vehicle occupancy factor \cite{united_states_department_of_transportation_federal_highway_policy__guidance_center_average_2018}. Concurrently, the maximum air-side capacity was determined by modeling continuous operations at full throughput, assuming a 90-second interval between takeoffs or landings, with the largest suitable aircraft for each facility. The resulting capacity estimates for the airport network are illustrated in Figure \ref{fig:road_airport_capacity}.  

\begin{table}[]
    \small
    \centering
    \caption{Runway Dimensions at Different Airports}
    \label{tab:airports}
    \begin{tabularx}{\textwidth}{@{}>{\raggedright\arraybackslash}p{3cm} *{4}{X} @{}} 
    \hline
    \textbf{Airport Name} & \textbf{Runway 1 Dimensions} & \textbf{Runway 2 Dimensions} & \textbf{Runway 3 Dimensions} & \textbf{Runway 4 Dimensions} \\
    \hline
    Watsonville Muni Airport-WVI & 4502 x 149 ft / 1372 x 45 m & 3998 x 98 ft / 1219 x 30 m &  &  \\
    San Martin Airport & 3095 x 75 ft / 943 x 23 m &  &  &  \\
    Sonoma Valley Airport (0Q3) & 2700 x 45 ft / 823 x 14 m & 1513 x 50 ft / 461 x 15 m &  &  \\
    Sonoma Skypark & 2490 x 40 ft / 759 x 12 m &  &  &  \\
    Rio Vista Municipal Airport & 4199 x 75 ft / 1280 x 23 m & 2199 x 60 ft / 670 x 18 m &  &  \\
    Petaluma Municipal Airport & 3600 x 75 ft / 1097 x 23 m &  &  &  \\
    Nut Tree Airport & 4700 x 75 ft / 1433 x 23 m &  &  &  \\
    Hollister Municipal Airport & 6350 x 100 ft / 1935 x 30 m & 3149 x 100 ft / 960 x 30 m &  &  \\
    Healdsburg Municipal Airport & 2652 x 60 ft / 808 x 18 m &  &  &  \\
    Half Moon Bay Airport & 5000 x 150 ft / 1524 x 46 m &  &  &  \\
    Gnoss Field Airport & 3303 x 75 ft / 1007 x 23 m &  &  &  \\
    Cloverdale Airport & 2909 x 60 ft / 887 x 18 m &  &  &  \\
    Byron Airport & 4500 x 100 ft / 1372 x 30 m & 3000 x 75 ft / 914 x 23 m &  &  \\
    Angwin-Parrett Field & 3217 x 50 ft / 981 x 15 m &  &  &  \\
    San Carlos Airport & 2621 x 75 ft / 799 x 23 m &  &  &  \\
    Reid-Hillview County Airport & 3100 x 75 ft / 945 x 23 m & 3099 x 75 ft / 945 x 23 m &  &  \\
    Palo Alto Airport & 2441 x 70 ft / 744 x 21 m &  &  &  \\
    Napa County Airport & 5930 x 150 ft / 1807 x 46 m & 5008 x 150 ft / 1526 x 46 m & 2510 x 75 ft / 765 x 23 m &  \\
    Livermore Municipal Airport & 5253 x 100 ft / 1601 x 30 m & 2699 x 75 ft / 823 x 23 m &  &  \\
    Hayward Executive Airport & 5694 x 150 ft / 1736 x 46 m & 3108 x 75 ft / 947 x 23 m &  &  \\
    Buchanan Field Airport & 5001 x 150 ft / 1524 x 46 m & 4602 x 150 ft / 1403 x 46 m & 2798 x 75 ft / 853 x 23 m & 2770 x 75 ft / 844 x 23 m \\
    \hline
    \end{tabularx}
    \end{table}

\subsection{Aircraft's Flight Simulation}
For this study, individual flights are modeled using a detailed five-phase profile based on FAA standards \cite{federal_aviation_administration_faa_2025}. The specific parameters for each phase are defined as follows:

\begin{enumerate}
    \item \textbf{Take-off:} IAS: 85 KTS 
    \item \textbf{Climb:} IAS: 115 KTS, ROC: 800 ft/min
    \item \textbf{Cruise:} IAS: 160 KTS, ALT: 5000 ft
    \item \textbf{Descent \& Approach:} IAS: 120 KTS, ROD: 500 ft/min
    \item \textbf{Landing:} IAS: 75 KTS
\end{enumerate}

Given that the largest distance between two airports in our study is 156 miles, we assume the above flight profile at a cruise altitude of 5000 ft, which lies within the performance envelope of the aircraft.

We further simplify this for simulation by calculating a single averaged runway-to-runway speed and time. The distance traveled and time taken in each phase of the flight is calculated as shown:

\subsubsection{Climb Phase}

\textbf{Time for Climb:}
\[
t_{\text{climb}} = \frac{\text{Cruise Altitude}}{\text{Rate of Climb (ROC)}} = \frac{5000}{800} = 6.25 \, \text{minutes}
\]

\noindent\textbf{Distance during Climb:}
\[
d_{\text{climb}} = \text{Climb IAS} \times \frac{t_{\text{climb}}}{60} = 115 \times \frac{6.25}{60}
\]
\[
d_{\text{climb}} = 115 \times 0.1042 \approx 11.98 \, \text{nautical miles (nm)} \approx 12.79 \, \text{miles}
\]

\subsubsection{Descent Phase}

\textbf{Time for Descent:}
\[
t_{\text{descent}} = \frac{\text{Cruise Altitude}}{\text{Rate of Descent (ROD)}} = \frac{5000}{500} = 10 \, \text{minutes}
\]

\noindent\textbf{Distance during Descent:}
\[
d_{\text{descent}} = \text{Descent IAS} \times \frac{t_{\text{descent}}}{60} = 120 \times \frac{10}{60}
\]
\[
d_{\text{descent}} = 120 \times 0.1667 \approx 20 \, \text{nm} \approx 23.02 \, \text{miles}
\]

\subsubsection{Travel Time for Aero Trips}

\textbf{Travel Time:}
\[
\text{Time} =
\begin{cases} 
\frac{\text{Distance}}{\frac{v_1 + v_2}{2}}, & \text{if } \text{Distance} \leq d_1 + d_2, \\[10pt]
t_{\text{climb}} + t_{\text{descent}} + d_3 / v_3, & \text{if } \text{Distance} > d_1 + d_2.
\end{cases}
\]

where:
\[
\begin{aligned}
d_1 &= d_{\text{climb}} = 12.79, \quad v_1 = 115 \, \text{mph}, \\
d_2 &= d_{\text{descent}} = 23.02, \quad v_2 = 120 \, \text{mph}, \\
d_3 &= \text{Distance} - d_1 - d_2, \quad v_3 = 160 \, \text{mph}.
\end{aligned}
\]

To determine the minimum operational interval for each airport runway, we apply the constraints outlined in FAA Order 7110.65 \cite{federal_aviation_administration_faa_2025}. This interval is a function of two primary components: Runway Occupancy Time (ROT) and the Aircraft Separation Requirement (ASR). For eVTOLs, we assume an ROT of approximately 30 seconds, with a slightly longer ROT for general aviation aircraft like Cessna 208 Caravan. The ASR, which ensures safety between successive aircraft, is set at 60 seconds under Visual Flight Rules (VFR) for small aircraft following another. Therefore, the Minimum Time Interval (T) is the sum of these values, which yields 90 seconds ($30s ROT + 60s ASR$). Consequently, the maximum theoretical capacity for a single runway is 40 operations per hour ($3600s / 90s$), where an operation can be either a take-off or landing.

\section{Result}
\label{sec:Result}
\subsection{Simulation of Heterogeneous Fleet Operation}
The equilibrium algorithm is executed until the convergence is achieved to get the time-saving situation for the UAM network in the Bay Area, under a standard 180-second take-off separation operation procedure. In total, our analysis indicates that approximately 230,000 travelers, representing 1.3\% of all regional trips, would experience time savings of more than 20 minutes. To accommodate this level of demand, the UAM network would need to conduct nearly 5,000 daily flights using a heterogeneous fleet of 241 aircraft. For perspective, if this service were operated exclusively with a smaller aircraft like the Cessna 172, the required number of flights would increase dramatically to 80,000. This flight volume is over 70 times greater than the combined daily operations of the three major Bay Area airports (SFO, SJC, and OAK, which total around 1,100 flights). In contrast to the high-volume operational model projected by Uber Elevate, which required 300 aircraft to conduct 12,000 daily flights for 48,000 passengers, our heterogeneous fleet strategy demonstrates greater efficiency by serving more passengers with a smaller fleet and fewer flights.

\begin{figure}[!ht]
  \centering
  \includegraphics[width=0.6\textwidth]{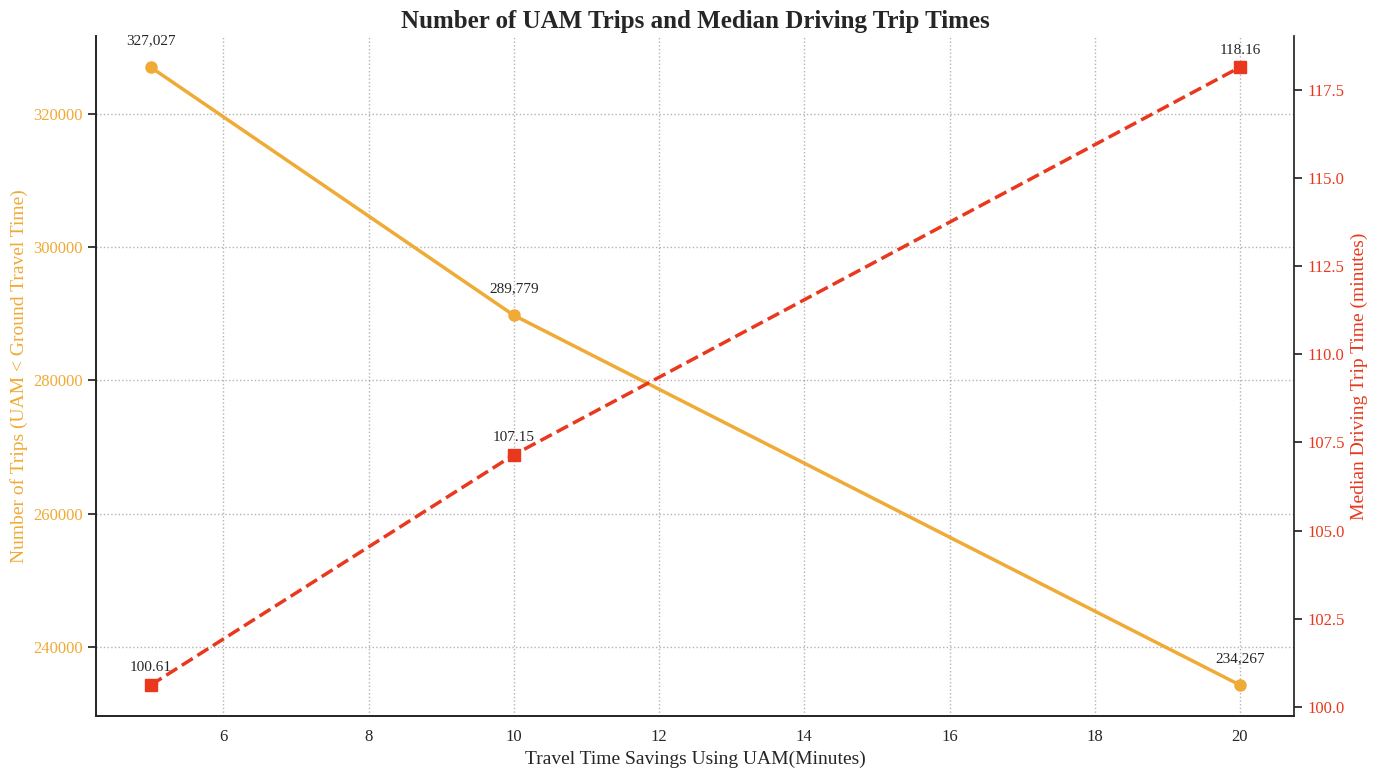}
  \caption{Number of Benefited Trips and Median Driving Time Based on Different Time Savings}\label{fig:NO_UAM_Trips_and_Mean_Driving_time}
\end{figure}

\begin{figure}[!ht]
  \centering
  \includegraphics[width=0.6\textwidth]{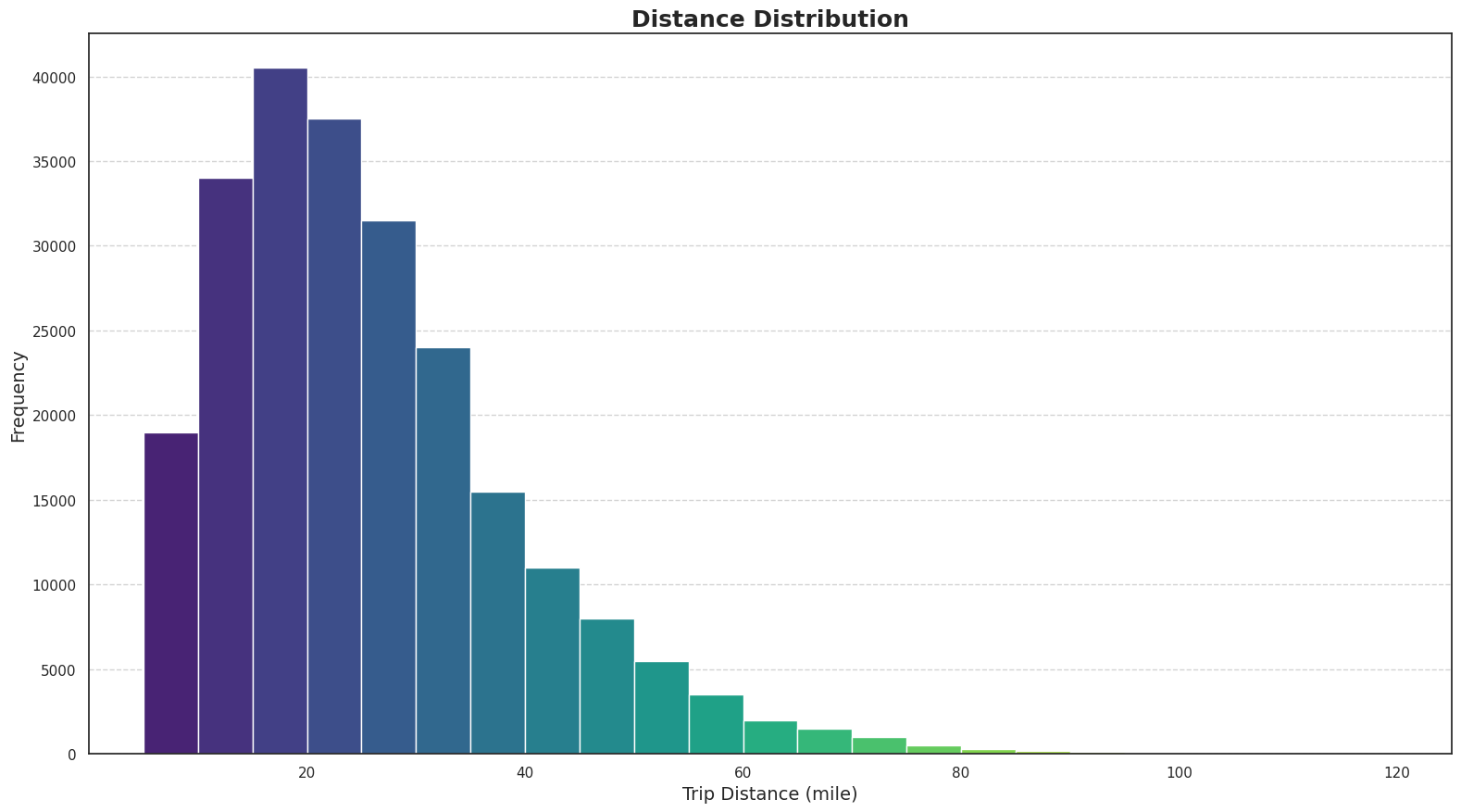}
  \caption{Distance Distribution for UAM Trips}\label{fig:Dist_Distribution}
\end{figure}

Figure \ref{fig:NO_UAM_Trips_and_Mean_Driving_time} plots two key metrics as a function of the travel time saved using UAM: the number of trips that benefit from the UAM network and the median driving time. On the one hand, as the median driving time increases, the time saved by the UAM network also increases nearly linearly. Starting with a zero-minute time savings threshold, there are 327,027 trips where the UAM network is faster than ground transportation. As the threshold increases to 10 and 20 minutes, the number of qualifying UAM trips decreases to 289,779 and 234,267 respectively. Therefore, as the threshold for what qualifies as significant time savings increases, the number of trips that meet this criterion decreases. However, travel time savings using UAM increase nearly linearly as the median driving time increases. Starting with a zero-minute time savings threshold, the median driving time is around 100 minutes. As the threshold increases to 10 and 20 minutes, the median driving time increases to 107 and 118 minutes, respectively. Figure \ref{fig:Dist_Distribution} shows the distance distribution for UAM trips. The concentration of trips in the 20-40 mile range aligns with the UAM’s sweet spot, demonstrating its potential as a middle ground between short car trips and long flights.

\subsection{Combination of Heterogeneous Fleet}
\begin{figure}[!ht]
  \centering
  \includegraphics[width=0.6\textwidth]{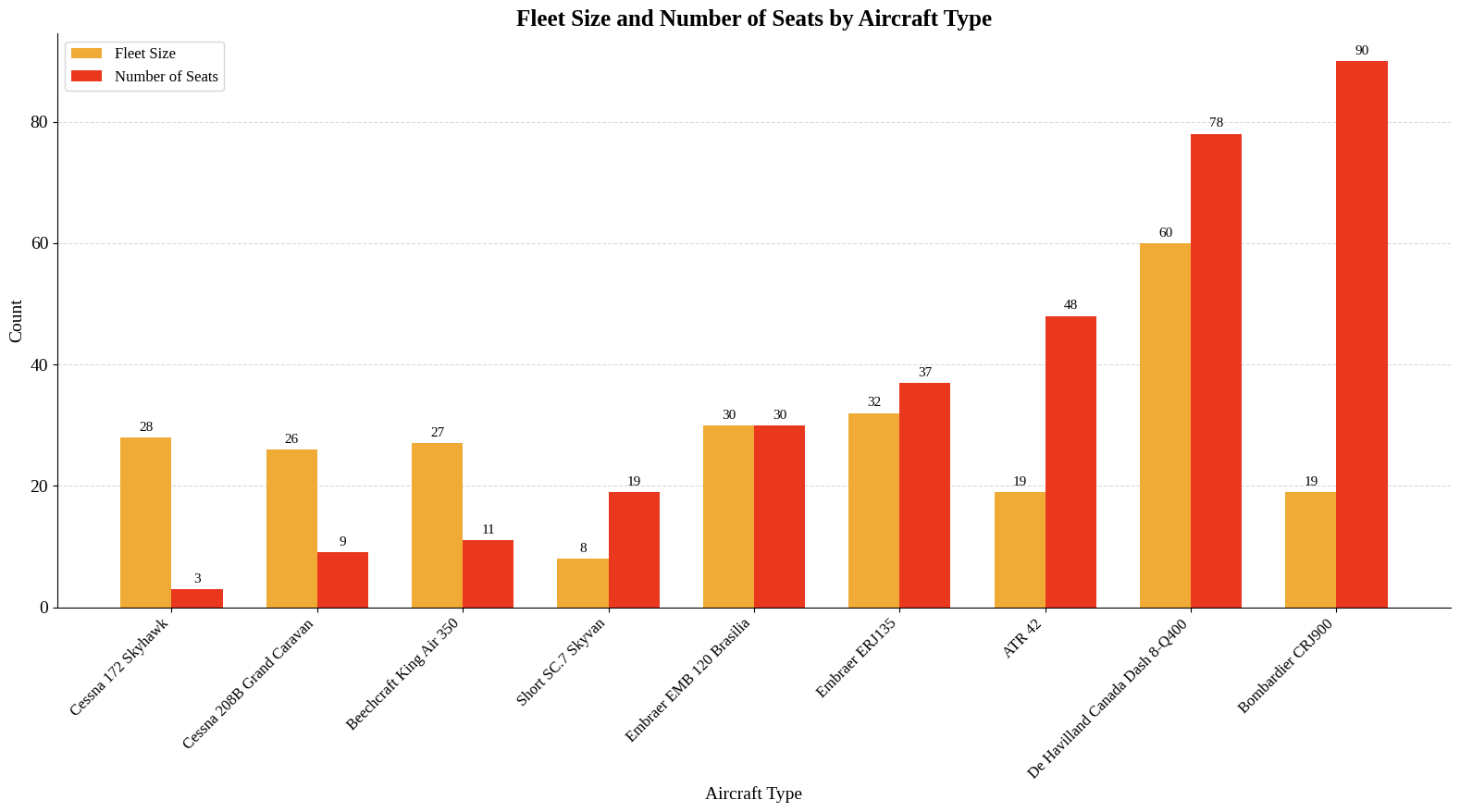}
  \caption{Fleet Size and Number of Seats by Aircraft Type}\label{fig:Hetero_Fleet_Size}
\end{figure}

Effective operation of a heterogeneous fleet requires precise allocation of aircraft type to maximize efficiency and control costs. Figure \ref{fig:Hetero_Fleet_Size} illustrates our calculation of the most efficient fleet composition in the Bay Area case. For low-demand periods, smaller aircraft like the Cessna 172 and Cessna 208 are employed to prevent capacity waste. During high-demand hours, larger regional jets, such as the widely used Dash 8-Q400 and CRJ900, are deployed to manage peak passenger loads and minimize waiting times.

\begin{figure}[!ht]
  \centering
  \includegraphics[width=0.6\textwidth]{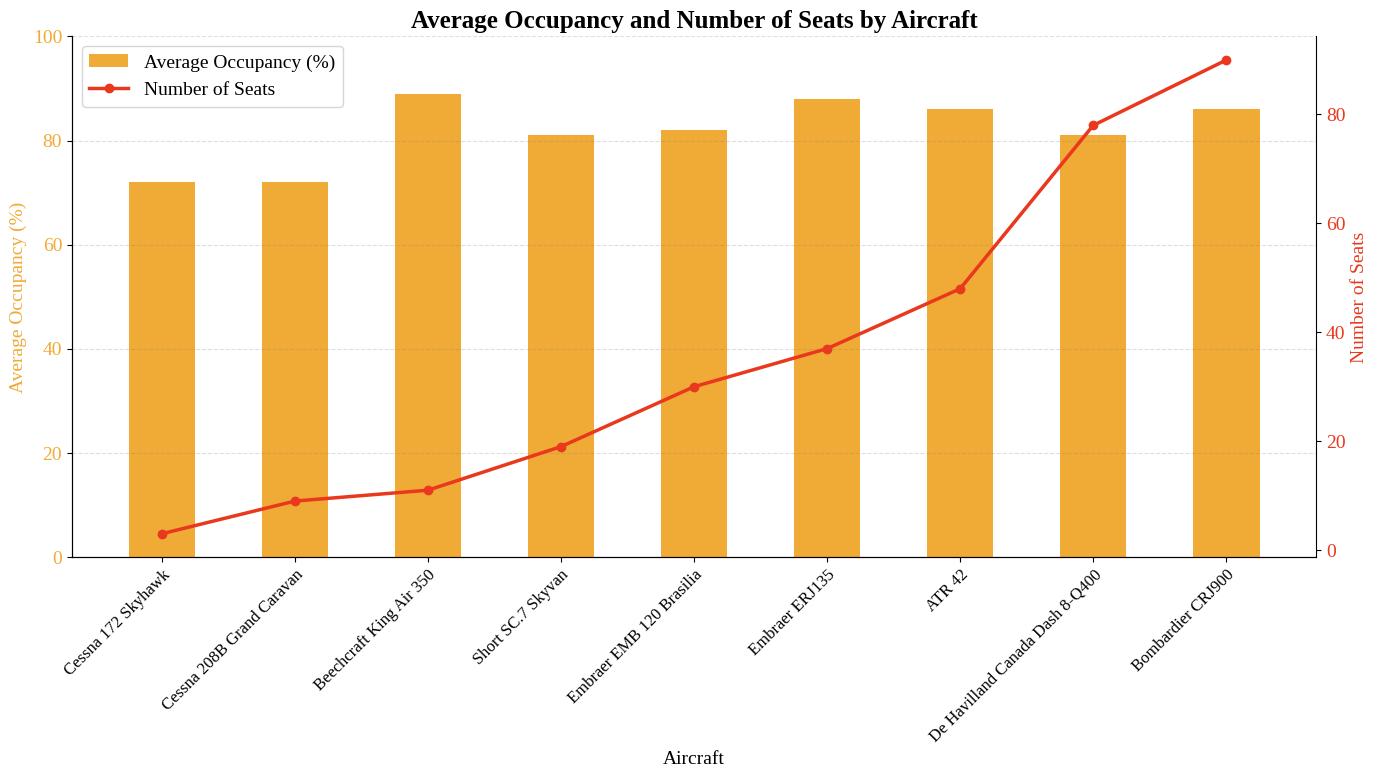}
  \caption{Average Occupancy and number of Seats by Aircraft}\label{fig:Occupancy}
\end{figure}

Figure \ref{fig:Occupancy} indicates the varying daily occupancy rates across different aircraft types. Generally, medium-sized and large aircraft exhibit higher average occupancy, while smaller aircraft typically show lower rates, with the exception of the Beechcraft King Air 350. This pattern is logical, as larger aircraft are deployed to accommodate greater passenger volumes. With a three-minute take-off interval, larger aircraft achieve higher occupancy when passenger crowds are substantial. Conversely, smaller aircraft maintain these intervals even during low passenger flow, leading to their comparatively lower average occupancy. Notably, all aircraft types maintain an average occupancy exceeding 70\%. 

\subsection{Temporal Scheduling Strategy}
\begin{figure}[!ht]
  \centering
  \includegraphics[width=0.6\textwidth]{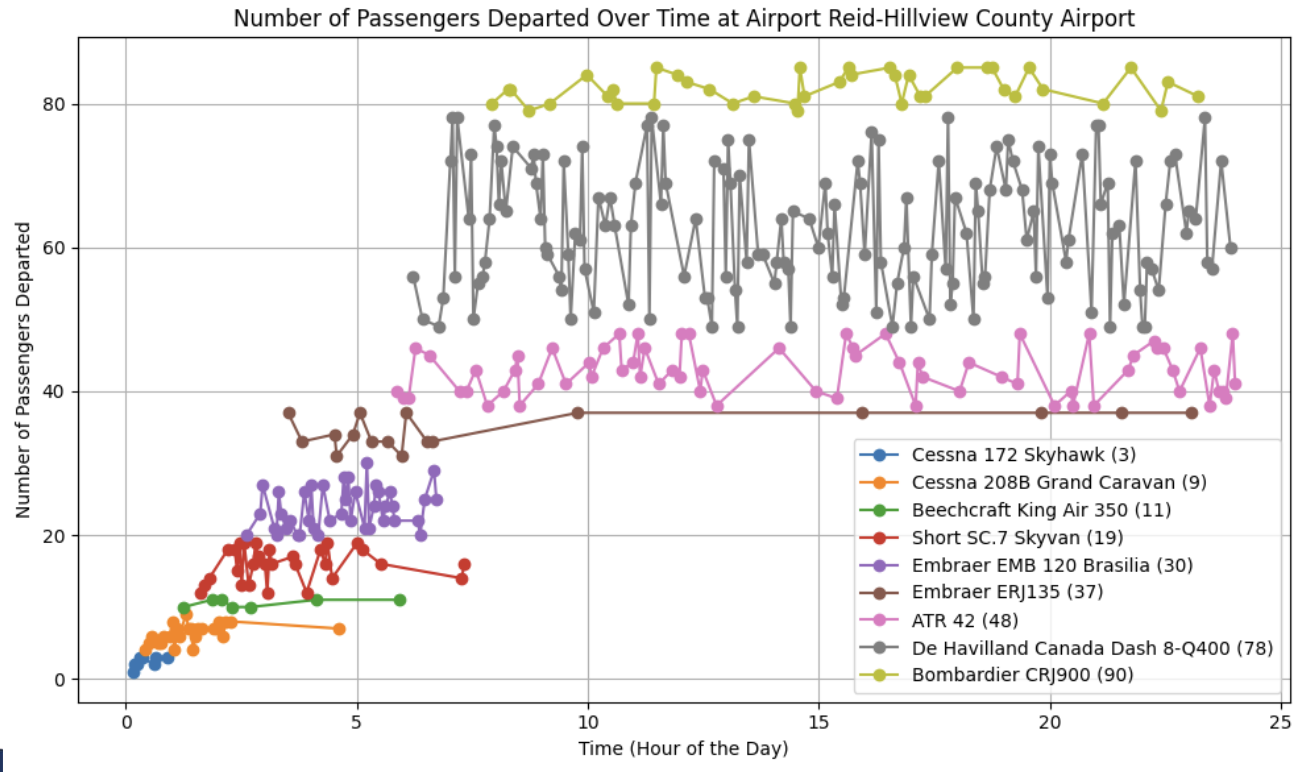}
  \caption{Demand Variations by Time of Day (Seat Capacity per Aircraft Type Shown in Parentheses))}\label{fig:NoPAX_departure_RHV}
\end{figure}

The temporal variation in UAM demand throughout the day necessitates strategic aircraft deployment to efficiently meet \ref{fig:NoPAX_departure_RHV} illustrates the estimated temporal distribution of each aircraft type at Reid-Hillview County Airport (RHV), where each distinct color represents a specific aircraft type and each dot signifies a take-off event.

During the early morning hours (0-5 AM), demand remains low, leading to the exclusive use of smaller aircraft such as the Cessna 172, Cessna 208, King Air 350, and SC.7 Skyvan. In this period, the primary constraint is often the low demand for land-side ground transportation to the airports. However, demand significantly increases after 7 AM, aligning with typical commuting patterns. Consequently, larger aircraft, including the ATR 42, Dash 8-Q400, and CRJ900, are deployed to accommodate this surge. While airports ideally operate at full capacity during peak times, our simulation indicates continued utilization of Dash 8-Q400, ATR 42, and even the 37-seats ERJ135, which are not the largest in our fleet, even during these high-demand periods. This observation suggests that factors beyond air-side flight capacity might be limiting full utilization of the largest aircraft.

\begin{figure}[!ht]
    \centering 
    \begin{minipage}[t]{0.48\textwidth}
        \centering
        \includegraphics[width=\linewidth]{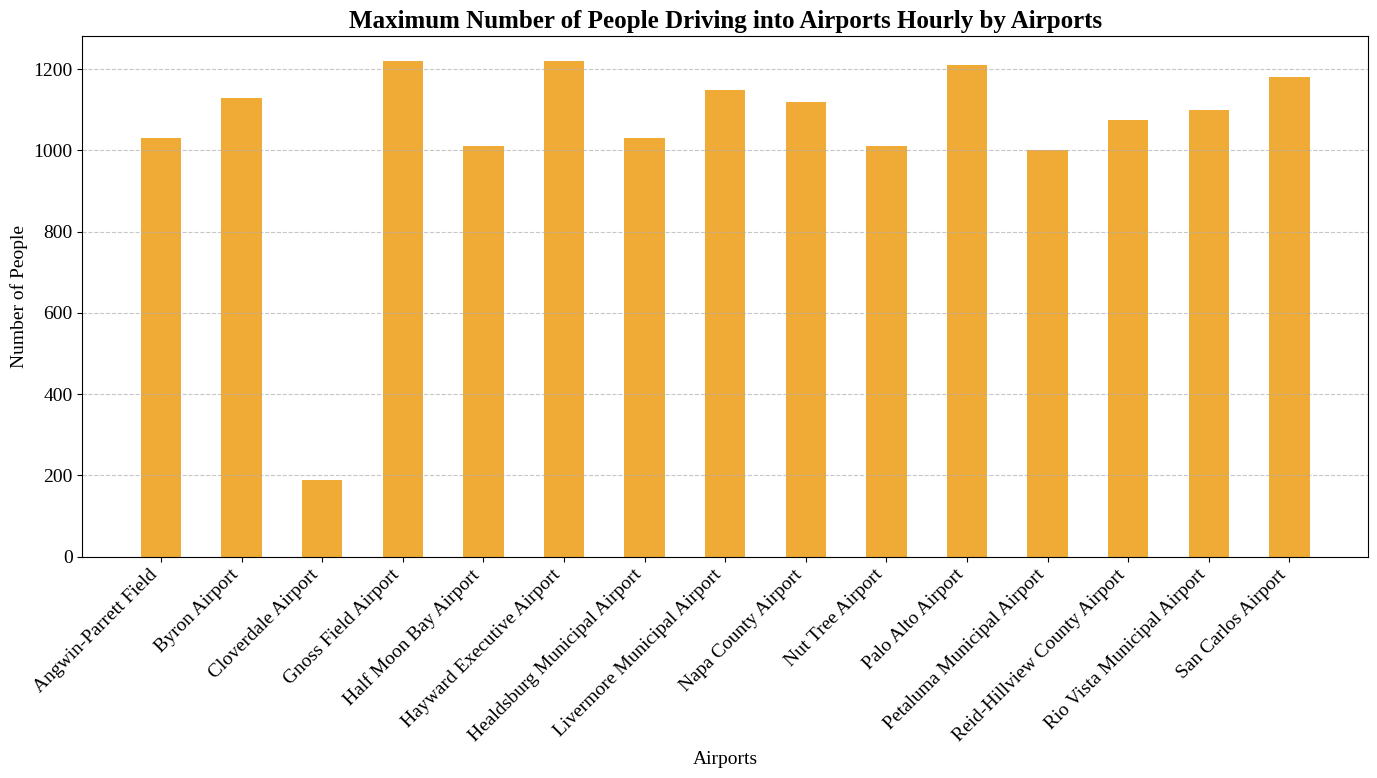} 
        \subcaption{} 
        \label{fig:CP_a} 
    \end{minipage}
    \hfill 
    \begin{minipage}[t]{0.48\textwidth}
        \centering
        \includegraphics[width=\linewidth]{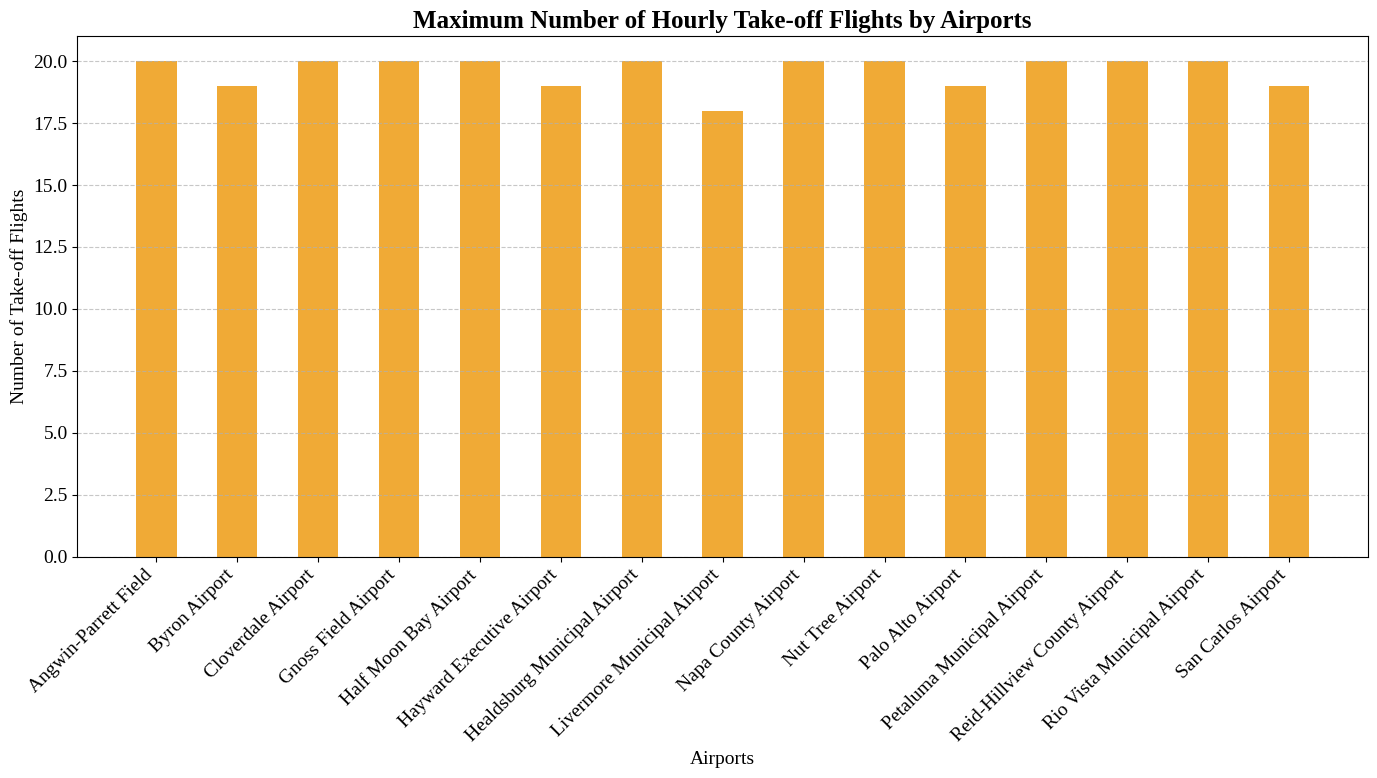} 
        \subcaption{} 
        \label{fig:CP_b} 
    \end{minipage}
    \caption{Comparison between Land-side Capacity(a) and Air-side Capacity(b)} 
    \label{fig:Comp_Landside_Airside_CP} 
\end{figure}

To pinpoint these operational bottlenecks, we compare the actual maximum passenger flow to airports via ground transportation with the maximum potential air-side flight operations at each airport, as presented in Figure \ref{fig:Comp_Landside_Airside_CP}. This analysis reveals that during peak hours, the air-side infrastructure is not consistently operating at its maximum capacity. Instead, the primary constraint often resides in the limited demand for UAM in the land-side ground transportation. A clear example is Cloverdale Airport, situated in the valley area on the northwestern side of the Bay Area. This small airport receives only ~200 ground-transported passengers during peak hours (Figure \ref{fig:CP_a}), despite having air-side operational capacity comparable to larger airports (Figure \ref{fig:CP_b}). Figure \ref{fig:road_airport_capacity} further demonstrates that while ground transportation could theoretically handle 5,814 passengers/hour, current UAM demand remains low. 

From our analysis, we observe a clear mismatch between UAM's available capacity and its actual usage. This gap demonstrates that current limitations stem primarily from insufficient demand rather than infrastructure constraints. These findings lead us to conclude that a one-size-fits-all approach to UAM deployment would be inefficient. Instead, we recommend prioritizing airport investments based on detailed demand assessments for each region.

\section{Conclusion}
\label{sec:Conclusion}
In this study, we integrate a heterogeneous fleet UAM system into the existing ground transportation network of the San Francisco Bay Area. Our analysis demonstrates that repurposing underutilized regional airports enables more than 230,000 travelers (1. 3\% of the 17.8 million daily origin-destination trips in the region) to achieve time savings of more than 20 minutes through UAM. To realize this potential, we develop an operational strategy that dynamically deploys smaller aircraft during off-peak hours and larger aircraft during peak demand. Using LPSim, a GPU-accelerated scheduling algorithm, we optimize the fleet composition to serve these passengers with only 241 aircraft that conduct 5,000 daily flights at 180-second take-off intervals.

Further analysis of simulation results confirms that the LPSim framework effectively models UAM-ground system interactions, enabling strategic planning (demand forecasting, fleet sizing, and network design). This iterative tool-based approach validates the role of UAM as a mid-distance solution (20-40 miles) within multimodal transport networks.

However, two critical challenges require future research. First, our simulation focuses on operational feasibility without evaluating economic viability, which requires a cost-benefit analysis. Second, the proposed 5,000 daily flights represent a four-fold increase over current air traffic in the Bay Area, posing substantial challenges for air traffic control and airspace management despite fleet heterogeneity.

\newpage

\bibliographystyle{trb}
\bibliography{trb_template}
\end{document}